\documentclass[twocolumn,aps,10pt,prl]{revtex4-2}

\usepackage[usenames]{xcolor}
\usepackage[english]{babel}
\usepackage[small]{caption}
\usepackage{verbatim}
\usepackage{multirow}
\usepackage{relsize}
\usepackage{amsmath,amssymb,amsbsy,amstext,amsfonts}
\usepackage{epsfig,bbm,simplewick}
\usepackage{mathtools}
\usepackage{enumitem}
\usepackage{csquotes}
\usepackage{tensor}
\usepackage{fix-cm}
\usepackage{booktabs}
\usepackage{graphicx}
\usepackage{bm}
\usepackage{subfigure}
\usepackage{float}
\usepackage[explicit]{titlesec}
\usepackage{enumitem}
\usepackage{setspace}
\usepackage{blindtext}
\usepackage{textpos}
\usepackage{etoolbox}
\usepackage[linktocpage=true]{hyperref}
\usepackage{anyfontsize}
\usepackage{listings,xfp}
\usepackage[dotinlabels]{titletoc}
\usepackage[normalem]{ulem}
\usepackage{esvect}
\DeclareMathAlphabet\mathbfcal{OMS}{cmsy}{b}{n}
\usepackage{tikz}
\usepackage{tikz-feynman}
\usetikzlibrary{patterns.meta}

\newcommand\numberthis{\addtocounter{equation}{1}\tag{\theequation}}
\hypersetup{colorlinks=true,linkcolor=blue,citecolor=blue,urlcolor=blue}
\allowdisplaybreaks

\begin{document}

\begin{flushleft}
\footnotesize DESY-24-177
\end{flushleft}

\phantom{solution}

\title{A consistency relation for induced \\ gravitational wave anisotropies}

\author{Juli\'an Rey}\email{julian.rey@desy.de}
\affiliation{Deutsches Elektronen-Synchrotron DESY, Notkestr.\,85, 22607 Hamburg, Germany}

\begin{abstract}
We show that the anisotropies in the spectrum of gravitational waves induced by scalar modes after the end of inflation in canonical, single-field models are completely determined by the tilt of the scalar and tensor power spectra. The latter contains information about anisotropies produced due to the propagation of the tensor modes in an inhomogeneous Universe, whereas the former represents the anisotropies generated at the time of production and arise only when non-Gaussian corrections to the angular power spectrum are considered. Our proof takes into account all scalar interactions in the cubic inflaton Lagrangian.
\end{abstract}

\maketitle

\section{INTRODUCTION}

\noindent Upon expanding Einstein's equations to second order in perturbations, one finds that gravitational waves are sourced by terms quadratic in scalar modes \cite{Tomita:1967wkp,Matarrese:1992rp,Carbone:2004iv,Ananda:2006af,Baumann:2007zm}. The physics of these scalar-induced gravitational waves has been thoroughly explored in recent years, see e.g.\,\cite{Domenech:2021ztg} for a pedagogical review. This mechanism of gravitational wave production becomes particularly relevant in models which feature an enhancement in the scalar power spectrum. For instance, in the context of single-field inflation, an inflection point in the potential can lead to a phase of ultra-slow-roll in which the field slows down and the power spectrum develops a sharp feature, which then translates into a peak in the frequency spectrum of gravitational waves. This class of models is motivated by the fact that an enhanced scalar spectrum leads to large density fluctuations which later collapse into primordial black holes, compact objects that could account for the entirety of dark matter.

Other cosmological sources (such as phase transitions or topological defects) can also produce a peak in the spectrum, however, and a key question that remains open is whether one can use other observables to distinguish between different signals. One possibility is to use anisotropies in the gravitational wave background. Astrophysical sources typically produce anisotropies that are much larger in amplitude than those due to cosmological sources, making them a key observable in determining the origin of the signal. Significant efforts have been devoted recently to search for these anisotropies in Pulsar Timing Arrays \cite{NANOGrav:2023gor,NANOGrav:2023tcn}, but measuring them is a difficult task \footnote{For instance, even an isotropic background could produce large anisotropies due to interference between different sources \cite{Konstandin:2024fyo}.}. Scalar-induced gravitational waves represent one of the main candidates for new physics in this context \cite{NANOGrav:2023hvm,Franciolini:2023pbf} and thus understanding their anisotropies is of paramount importance for these analyses.

Since gravitational waves are generated locally (once the large scalar fluctuations that source them re-enter the horizon after inflation), one would expect distant patches in the sky to be essentially uncorrelated, and therefore for the corresponding anisotropies to be volume-suppressed. Although generally speaking this argument is correct, non-Gaussianities provide a way out of this issue. For instance, anisotropies induced by non-Gaussianities of the local kind (that is, when the curvature perturbation $\mathcal{R}$ can be written as $\mathcal{R}= \mathcal{R}_{\rm G}+f_{\rm NL}\mathcal{R}_{\rm G}^2$ for some Gaussian variable $\mathcal{R}_{\rm G}$) were considered in \cite{Bartolo:2019oiq,Bartolo:2019zvb,Bartolo:2019yeu,Dimastrogiovanni:2022afr,Li:2023qua,Li:2023xtl}, where it was shown that, since the product $\mathcal{R}_{\rm G}^2$ becomes a convolution in Fourier space, it is possible for two short-wavelength modes to conspire and create a long-wavelength one, effectively correlating distant patches \footnote{The non-Gaussian correction to the gravitational wave frequency spectrum have also been considered in \cite{Garcia-Bellido:2017aan,Cai:2018dig,Unal:2018yaa,Atal:2021jyo,Adshead:2021hnm,Ragavendra:2021qdu,Abe:2022xur,Papanikolaou:2024kjb,Perna:2024ehx}.}. As shown in \cite{Ruiz:2024weh}, the argument also holds when other types of non-Gaussianity (e.g.\,those that arise due to the self-interactions of the inflaton) are considered.

The goal of this work is to show that, in canonical, single-field models of inflation, anisotropies are completely determined by the tilt of the scalar and tensor spectrum on small scales. Concretely, let us anticipate that we find the coefficients $C_\ell$ for the angular power spectrum of the gravitational wave density contrast $\delta_{\rm GW}$ are given by \cite{Bartolo:2019zvb,Li:2023xtl}
\begin{equation}
C_\ell(q)
=
\frac{2\pi\mathcal{P}^{\rm L}_\mathcal{R}}{\ell(\ell+1)}
\bigg\{
\frac{\Omega_{\rm NG}(q)}{\Omega_{\rm GW}(q)}
+\frac{3}{5}\bigg[4-\frac{\partial\log\Omega_{\rm GW}(q)}{\partial\log q}\bigg]
\bigg\}^2.
\label{eq:anisotropies}
\end{equation}
where $\Omega_{\rm GW}$ is the induced gravitational wave energy density, $\mathcal{P}^{\rm L}_\mathcal{R}\sim 10^{-9}$ is the long-wavelength scalar power spectrum on CMB scales, and the term
\begin{align*}
&\Omega_{\rm NG}(q)
=
-\frac{1}{12}\int\frac{d^3p}{(2\pi)^3}
\frac{1}{2\pi^2q}
\Big[{\bm p}\cdot{\bm e}^s({\bm q})\cdot{\bm p}\Big]^2
\frac{(2\pi^2)^2}{p^3|{\bm q}-{\bm p}|^3}
\\&\;
\frac{q^2}{a^2H^2}\overline{I_q(p,|{\bm q}-{\bm p}|)^2}
\mathcal{P}_\mathcal{R}(|{\bm q}-{\bm p}|)
\mathcal{P}_\mathcal{R}(p)
\frac{d\log\mathcal{P}_\mathcal{R}(p)}{d\log p}
\numberthis
\label{eq:omega-ng-main}
\end{align*}
is identical to the usual expression for the scalar-induced $\Omega_{\rm GW}$ (see the following section for our notation and conventions), but where the integrand is modulated by the tilt of the scalar spectrum on small scales. The above equation is the main result of this paper and shows that, in single-field models of inflation, the non-Gaussian corrections to the angular power spectrum only appear in one specific combination, namely, the scalar bispectrum in the squeezed limit, which can then be related to the tilt of the scalar spectrum by using Maldacena's consistency relation \cite{Maldacena:2002vr,Creminelli:2004yq}. Previous works studying non-Gaussianties of the local kind can therefore be thought of as particular cases of this equation (if we restrict our attention to single-field models, for which the consistency relation in \cite{Maldacena:2002vr,Creminelli:2004yq} holds). This relation was first discussed in \cite{Ruiz:2024weh}, where non-Gaussianities induced by the inflaton self-interactions were considered. Here we show that this is not an accidental result, and in fact is a general relation which we obtain by considering the full single-field cubic Lagrangian in our proof.

\section{ANISOTROPIES}

\noindent We now provide a brief summary of the derivation of the expression for the angular power spectrum. The full calculation can be found with our conventions e.g.\,in \cite{Ruiz:2024weh}. We work in natural units and set the reduced Planck mass $M_p$ to unity.

At leading order in perturbations, scalar and tensor modes evolve independently. Once Einstein's equations are expanded to second order, however, we find that gravitational waves are sourced by terms quadratic in scalar modes. In Fourier space, the transverse-traceless tensor modes of the metric can be written as
\begin{equation}
h_{ij}({\bm x})
=
\int\frac{d^3k}{(2\pi)^3}
e^s_{ij}({\bm k})h_{\bm k}^s
e^{i{\bm k}\cdot{\bm x}},
\label{eq:tensor_fourier}
\end{equation}
where the index $s=+,\times$ refers to the two tensor polarizations, and the quantities $e^s_{ij}({\bm k})$ are symmetric, transverse-traceless tensors satisfying $k^i e_{ij}^s({\bm k})=0$ and $\delta^{ij}e_{ij}^s({\bm k})=0$. The equation of motion for $h_k^s$ is
\begin{equation}
h_k^{s\prime\prime}+2aHh_k^{s\prime}+k^2h_k^s=T_k^s,
\label{eq:tensor_eom}
\end{equation}
where primes denote derivatives with respect to conformal time $d\tau=dt/a$, and the source term $T_k^s$, which is quadratic in scalar perturbations, can be found, with our conventions, in \cite{Ruiz:2024weh}. This equation can be solved by Green's function techniques. The solution is
\begin{equation}
h_k^s(\tau)
=
\frac{1}{k^2}\int\frac{d^3p}{(2\pi)^3}
\Big[
{\bm p}\cdot{\bm e}^s({\bm k})\cdot{\bm p}
\Big]
\mathcal{R}_{\bm p}\mathcal{R}_{{\bm k}-{\bm p}}
I_k(\tau,p,|{\bm k}-{\bm p}|),
\label{eq:h_solution}
\end{equation}
where ${\bm p}\cdot{\bm e}^s({\bm k})\cdot{\bm p}
=p_ip_je^s_{ij}$ and $I_k(\tau,p,|{\bm k}-{\bm p}|)$ is given, in the radiation era, by
\begin{equation}
I_k(\tau,p,|{\bm k}-{\bm p}|)
=
\frac{4}{9}
\int kG_k(\tau,\tau')Q_k(\tau',p,|{\bm k}-{\bm p}|)kd\tau',
\end{equation}
where $G_k(\tau,\tau')$ is the Green's function of the equation of motion and $Q_k(\tau',p,|{\bm k}-{\bm p}|)$ is related to the source via
\begin{equation}
T_k^s(\tau)
=
\frac{4}{9}
\int\frac{d^3p}{(2\pi)^3}
\Big[
{\bm p}\cdot{\bm e}^s({\bm k})\cdot{\bm p}
\Big]
\mathcal{R}_{\bm p}\mathcal{R}_{{\bm k}-{\bm p}}
Q_k(\tau,p,|{\bm k}-{\bm p}|).
\end{equation}
The quantity of interest for us is the energy density of gravitational waves,
\begin{equation}
\label{eq:gw_energy}
\rho_{\rm GW}(\tau,{\bm x})=\frac{1}{16a^2}\overline{\partial_k h_{ij}({\bm x})\partial_k h_{ij}({\bm x})},
\end{equation}
where the bar denotes an average over time, which must be taken due to the stochastic nature of the signal \cite{maggiore}.

The phase-space distribution of a collection of gravitons $f(x^\mu,p^\mu)$ depends only on the comoving momentum $q=ap$ in the absence of scalar metric perturbations. When the particles propagate in an inhomogeneous Universe, however, this quantity is generically also a function of position $x^\mu$ and the direction of propagation is $\bm n={\bm p}/p$. The distribution obeys the free Boltzmann equation \footnote{Since we are interested in the primordial stochastic background, graviton emission can be treated as an initial condition. Self-interactions are Planck-suppressed and can therefore also be neglected, so the Boltzmann equation is unsourced.} and can then be written as the sum of a $q$-dependent background piece and a perturbation
\begin{equation}
f(\tau,{\bm x},q,\bm n)
=
\hat{f}(q)-q\frac{\partial\hat{f}}{\partial q}\Gamma(\tau,{\bm x},q,\bm n).
\label{eq:f_pert}
\end{equation}
We work in the Newtonian gauge assuming no anisotropic stress and ignore vector and tensor perturbations, so that the only relevant degree of freedom is the Newtonian potential $\psi$. By combining the geodesic equation for the gravitons with the free Boltzmann equation we can write a differential equation for $\Gamma$ with the following solution in Fourier space (see e.g.\,\cite{Ruiz:2024weh} for details)
\begin{equation}
e^{i{\bm k}\cdot{\bm n}\tau_\star}\Gamma_{\bm k}(\tau_\star)
=
e^{i{\bm k}\cdot{\bm n}\tau}\Big[\Gamma_{\bm k}(\tau)+\psi_{\bm k}(\tau)\Big],
\label{eq:bolt_sol}
\end{equation}
where we have neglected a monopole term and a term that encodes the gravitational wave equivalent of the integrated Sachs-Wolfe effect, which was estimated numerically in \cite{Bartolo:2019zvb} and found to be subdominant.

The gravitational wave energy density in eq.\,(\ref{eq:gw_energy}) is related to $f$ as follows
\begin{equation}
\rho_{\rm GW}({\bm x})= \frac{g}{a^4}\int \frac{d^3q}{(2\pi)^3} qf({\bm x},{\bm q}),
\label{eq:stat_energy}
\end{equation}
where the number of internal degrees of freedom is $g=2$ for gravitons. The density contrast is
\begin{equation}
\delta_{\rm GW}({\bm x},{\bm q})
=
\frac{f({\bm x},{\bm q})-\hat{f}(q)}{\hat{f}(q)}.
\label{eq:density_cont}
\end{equation}
The background contribution can be obtained by taking the ensemble average $\hat{f}(q)=\langle f({\bm x},{\bm q})\rangle\propto \Omega_{\rm GW}(q)/q^4$. Using (\ref{eq:f_pert}) and taking $\tau_\star$ as today in eq.\,(\ref{eq:bolt_sol}), we find the present density contrast
\begin{align*}
\delta_{\rm GW}(\tau_\star,{\bm x}_\star,{\bm q})
&=
\delta_{\rm GW}(\tau,{\bm x},{\bm q})
\\&
+\bigg[4-\frac{\partial\log\Omega_{\rm GW}(\tau,q)}{\partial\log q}\bigg]
\psi(\tau,{\bm x}),\numberthis
\label{eq:present_contrast}
\end{align*}
where ${\bm x}={\bm x}_\star+{\bm n}(\tau-\tau_\star)$.

We can decompose the density contrast into spherical harmonics,
\begin{equation}
\delta_{\rm GW}(\tau_\star,{\bm x}_\star,{\bm q})
=
\sum_{\ell m}\delta_{\ell m}(\tau_\star,{\bm x}_\star,q)Y_{\ell m}({\bm n}).
\end{equation}
The multipole moments $\delta_{\ell m}(\tau_\star,{\bm x}_\star,q)$ have the following two-point function
\begin{equation}
\langle
\delta_{\ell m}(\tau_\star,{\bm x}_\star,q)
\delta^{\star}_{\ell'm'}(\tau_\star,{\bm x}_\star,q)
\rangle
=
\delta_{\ell\ell'}\delta_{mm'}C_\ell(\tau_\star,{\bm x}_\star,q).
\end{equation}
After some manipulation \cite{Ruiz:2024weh} we obtain the following expression for the angular coefficients
\begin{equation}
C_\ell(\tau_\star,{\bm x}_\star,q)
=
4\pi\int\frac{dk}{k}
\mathcal{P}_{\delta_{\rm GW}}(\tau_\star,{\bm x}_\star,q,k)
j^2_\ell(kD),
\label{eq:spherical_bessel}
\end{equation}
where $j_\ell$ is the spherical Bessel function of the first kind, $D$ is a dimensionful quantity with units of distance and $\mathcal{P}_{\delta_{\rm GW}}$ is the dimensionless power spectrum of $\delta_{\rm GW}$.

\section{NON-GAUSSIANITIES}

\noindent In the in-in formalism, the expectation value of a Hermitian operator $Q$ is given, at second order, by the following expression \cite{Weinberg:2005vy}
\begin{align*}
&\langle Q(t)\rangle
=
\langle Q_I(t)\rangle
+
2{\rm Im}\bigg\{\int_{-\infty}^t dt'
\langle Q_I(t)\mathcal{H}(t')\rangle\bigg\}
\\&
+
2{\rm Re}\bigg\{\int_{-\infty}^t dt'\int_{t'}^t dt''
\langle \big[\mathcal{H}(t''),Q_I(t)\big]\mathcal{H}(t')\rangle\bigg\},
\label{eq:in-in-master}\numberthis
\end{align*}
where $Q_I$ is the operator in the interaction picture and $\mathcal{H}$ denotes the interaction Hamiltonian. It is convenient to express the different field contractions by using diagrams, as in \cite{Ballesteros:2024zdp,Li:2023xtl,Ruiz:2024weh}. Instead of using the inflaton perturbation $\delta\phi$, we work with the linearized curvature perturbation
\begin{equation}
\varphi \equiv -\frac{H}{\dot{\phi}}\delta\phi.
\end{equation}
The power spectrum of $\mathcal{R}$ is $\mathcal{P}_\mathcal{R}\equiv (k^3/2\pi^2)|\varphi_k|^2$ and $\varphi_k$ denotes the solution to the Mukhanov-Sasaki equation. In what follows we calculate the correlators $\langle\delta\phi^n\rangle$ in the $\delta\phi$ gauge and then perform a gauge transformation to obtain the correlators $\langle\mathcal{R}^n\rangle$. The transformation connecting both gauges is ($\cdot \equiv d/dt$) \cite{Maldacena:2002vr}
\begin{equation}
\mathcal{R}
=
\varphi-\frac{1}{2}\eta\varphi^2+\frac{1}{3}\bigg(\eta^2+\frac{\dot{\eta}}{2H}\bigg)\varphi^3+\mathcal{O}(\varphi^4).
\label{eq:gauge_trans}
\end{equation}
If $\varphi$ were Gaussian, this expression would correspond to local non-Gaussianity of the form $\mathcal{R}=\mathcal{R}_{\rm G}+f_{\rm NL}\mathcal{R}_{\rm G}^2+\cdots$ with $\mathcal{R}_{\rm G}=\varphi$ and nonlinearity parameter $f_{\rm NL}=-\eta/2$. The slow-roll parameters are
\begin{equation}
\epsilon=-\frac{\dot{H}}{H^2},\qquad \eta=-\frac{1}{2}\frac{\dot{\epsilon}}{H\epsilon}.
\end{equation}
In models of inflation featuring an ultra-slow-roll phase, modes around the peak of the spectrum freeze at the same time, during the subsequent constant-roll phase in which $\eta$ is a negative, constant $\mathcal{O}(1)$ number \cite{Ballesteros:2024pwn}. Motivated by these considerations, we denote the coefficient in the second term of (\ref{eq:gauge_trans}) by $f_{\rm NL}$ (even though $\varphi$ is not Gaussian in our setup), which we take as constant.

Fields are schematically represented by
\begin{equation}
\mathcal{R}
\;\sim\;
\begin{tikzpicture}[baseline={-2}]
    \draw [dashed] (0,0) -- (1,0);
\end{tikzpicture}\;,
\quad
\varphi
\;\sim\;
\begin{tikzpicture}[baseline={-2}]
    \draw (0,0) -- (1,0);
\end{tikzpicture}\;,
\quad
h
\;\sim\;
\begin{tikzpicture}[baseline={-2}]
    \draw [decorate,decoration={snake,amplitude=1.5pt,segment length=6pt}] (0,0) -- (1,0);
\end{tikzpicture}\;.
\end{equation}
In any correlator $\langle\mathcal{R}^n\rangle$ we must first turn each factor of $\mathcal{R}$ into powers of $\varphi$ using eq.\,(\ref{eq:gauge_trans}). We can think of the gauge transformation in terms of interaction vertices which we represent as black dots,
\begin{equation}
1
\;\sim\;
\begin{tikzpicture}[baseline={-2}]
    \draw [dashed] (0,0) -- (0.56,0);
    \draw (0.56,0) -- (1.0,0);
	\fill[black] (0.57,0) circle (2pt);
\end{tikzpicture}\;,
\quad
f_{\rm NL}
\;\sim\;
\begin{tikzpicture}[baseline={-2}]
    \draw [dashed] (0,0) -- (0.56,0);
    \draw (0.56,0) -- (0.81,0.5);
    \draw (0.56,0) -- (0.81,-0.5);
	\fill[black] (0.56,0) circle (2pt);
\end{tikzpicture}\;,
\quad
g_{\rm NL}
\;\sim\;
\begin{tikzpicture}[baseline={-2}]
    \draw [dashed] (0,0) -- (0.56,0);
    \draw (0.56,0) -- (0.81,0.5);
    \draw (0.56,0) -- (0.81,-0.5);
    \draw (0.56,0) -- (1.12,0);
	\fill[black] (0.56,0) circle (2pt);
\end{tikzpicture}\;.
\label{eq:gauge_ints}
\end{equation}
Once we have turned all dashed lines into solid lines, we can use the cubic Lagrangian in the $\delta\phi$ gauge to calculate the $\langle\delta\phi^n\rangle$ correlators. The cubic self-interactions of the inflaton are generically represented by white dots,
\begin{equation}
\varphi^3,\;\varphi\dot{\varphi}^2,\;{\rm etc.}
\;\sim\;
\begin{tikzpicture}[baseline={-2}]
    \draw (0,0) -- (0.56,0);
    \draw (0.56,0) -- (0.81,0.5);
    \draw (0.56,0) -- (0.81,-0.5);
	\fill[black] (0.56,0) circle (2.5pt);
	\fill[white] (0.56,0) circle (1.5pt);
\end{tikzpicture}\;.
\end{equation}
We represent the long-wavelength $\varphi$ on CMB scales with a double line
\begin{equation}
\varphi_{\rm L}
\;\sim\;
\begin{tikzpicture}
    \draw (0,0.54) -- (1,0.54);
    \draw (0,0.46) -- (1,0.46);
\end{tikzpicture}\;.
\label{eq:double_solid}
\end{equation}

The gravitational wave density contrast $\delta_{\rm GW}=\delta\rho_{\rm GW}/\rho_{\rm GW}$ is, diagrammatically,
\begin{equation}
\delta_{\rm GW}
\;\sim\;
\begin{tikzpicture}[baseline={-2}]
    \draw [decorate,decoration={snake,amplitude=1pt,segment length=6pt}] (0,0.035) -- (1,0.035);
    \draw [decorate,decoration={snake,amplitude=1pt,segment length=6pt}] (0,-0.035) -- (1,-0.035);
\end{tikzpicture}\;.
\end{equation}
The vertices $h\sim \mathcal{R}^2$ and $\delta_{\rm GW}\sim h^2$ are represented as one wavy line splitting into two dashed ones, and one double wavy line splitting into two single ones.

As shown in \cite{Bartolo:2019zvb,Bartolo:2019yeu,Ruiz:2024weh}, the largest contribution to the angular power spectrum is found by using cubic vertices to join the two halves of the following diagram with a single long-wavelength propagator $|\varphi_{\rm L}|^2$,
\begin{equation}
\langle\delta_{\rm GW}^2\rangle\;\sim\;
\begin{tikzpicture}[baseline={-2}]
    \draw [decorate,decoration={snake,amplitude=1pt,segment length=6pt}] (-0.5,0.035) -- (0.3,0.035);
    \draw [decorate,decoration={snake,amplitude=1pt,segment length=6pt}] (-0.5,-0.035) -- (0.3,-0.035);
    \draw [decorate,decoration={snake,amplitude=1.5pt,segment length=6pt}] (0.3,0) to[out=90,in=180] (1.1,0.7);
    \draw [decorate,decoration={snake,amplitude=1.5pt,segment length=6pt}] (0.3,0) to[out=-90,in=180] (1.1,-0.7);
    \draw [dashed] (1.1,+0.7) -- (1.45,+0.35);
    \draw [dashed] (1.1,+0.7) -- (0.75,+0.35);
    \draw [dashed] (1.1,-0.7) -- (1.45,-0.35);
    \draw [dashed] (1.1,-0.7) -- (0.75,-0.35);
    \draw (0.75,+0.35) -- (0.75,-0.35);
	\fill[black] (0.75,+0.35) circle (2pt);
	\fill[black] (0.75,-0.35) circle (2pt);
    \draw[color=black] (1.45,-0.35) rectangle (2.15,0.35);
    \fill[pattern=north east lines, pattern color=black] (1.45,-0.35) rectangle (2.15,0.35);
	\fill[white] (1.45,+0.35) circle (3pt);
	\fill[black] (1.45,+0.35) circle (2pt);
 	\fill[white] (2.15,+0.35) circle (3pt);
	\fill[black] (2.15,+0.35) circle (2pt);
  	\fill[white] (1.45,-0.35) circle (3pt);
	\fill[black] (1.45,-0.35) circle (2pt);
   	\fill[white] (2.15,-0.35) circle (3pt);
	\fill[black] (2.15,-0.35) circle (2pt);    
    \draw (2.85,+0.35) -- (2.85,-0.35);
	\fill[black] (2.85,+0.35) circle (2pt);
	\fill[black] (2.85,-0.35) circle (2pt);
    \draw [dashed] (2.5,+0.7) -- (2.15,+0.35);
    \draw [dashed] (2.5,+0.7) -- (2.85,+0.35);
    \draw [dashed] (2.5,-0.7) -- (2.15,-0.35);
    \draw [dashed] (2.5,-0.7) -- (2.85,-0.35);
    \draw [decorate,decoration={snake,amplitude=1.5pt,segment length=6pt}] (2.5,0.7) to[out=0,in=90] (3.3,0);
    \draw [decorate,decoration={snake,amplitude=1.5pt,segment length=6pt}] (2.5,-0.7) to[out=0,in=-90] (3.3,0);
    \draw [decorate,decoration={snake,amplitude=1pt,segment length=6pt}] (4.1,0.035) -- (3.3,0.035);
    \draw [decorate,decoration={snake,amplitude=1pt,segment length=6pt}] (4.1,-0.035) -- (3.3,-0.035);
\end{tikzpicture}\;.
\label{eq:master-diagram}
\end{equation}
In other words, the shaded box in this diagram must be replaced with all possible diagrams connecting both sides with only one double solid line, as in eq.\,(\ref{eq:double_solid}). The corresponding diagrams are
\begin{equation}
\begin{tikzpicture}[baseline={-2}]
	\fill[black] (0,+0.35) circle (2pt);
	\fill[black] (0.7,+0.35) circle (2pt);
	\fill[black] (0,-0.35) circle (2pt);
	\fill[black] (0.7,-0.35) circle (2pt);
    \draw (0,-0.35) -- (0,+0.35);
    \draw (0,+0.39) -- (0.7,+0.39);
    \draw (0,+0.31) -- (0.7,+0.31);
    \draw (0.7,+0.35) -- (0.7,-0.35);
\end{tikzpicture}
\quad+\quad
\begin{tikzpicture}[baseline={-2}]
	\fill[black] (0,+0.35) circle (2pt);
	\fill[black] (0.7,+0.35) circle (2pt);
	\fill[black] (0,-0.35) circle (2pt);
	\fill[black] (0.7,-0.35) circle (2pt);
    \draw (0,-0.35) -- (0.175,0);
    \draw (0,0.35) -- (0.175,0);
    \draw (0.175,0.04) -- (0.525,0.04);
    \draw (0.175,-0.04) -- (0.525,-0.04);
    \draw (0.525,0) -- (0.7,0.35);
    \draw (0.525,0) -- (0.7,-0.35);
	\fill[black] (0.175,0) circle (2.5pt);
	\fill[white] (0.175,0) circle (1.5pt);
	\fill[black] (0.525,0) circle (2.5pt);
	\fill[white] (0.525,0) circle (1.5pt);
\end{tikzpicture}
\quad+\quad
\begin{tikzpicture}[baseline={-2}]
	\fill[black] (0,+0.35) circle (2pt);
	\fill[black] (0.7,+0.35) circle (2pt);
	\fill[black] (0,-0.35) circle (2pt);
	\fill[black] (0.7,-0.35) circle (2pt);
    \draw (0,-0.35) -- (0,+0.35);
    \draw (0,+0.39) -- (0.35,0.04);
    \draw (0,+0.31) -- (0.35,-0.04);
    \draw (0.35,0) -- (0.7,-0.35);
    \draw (0.35,0) -- (0.7,0.35);
	\fill[black] (0.35,0) circle (2.5pt);
	\fill[white] (0.35,0) circle (1.5pt);
\end{tikzpicture}\;.
\label{eq:s_channel}
\end{equation}
There are many other possible ways of connecting both sides of this diagram (for instance, using a quartic interaction, or even at the Gaussian level by connecting the dashed lines together), but these either lead to volume-suppressed results or vanish once we take the propagator connecting both sides to be long-wavelength (see e.g.\,\cite{Ruiz:2024weh} for more details). We neglect first-order tensor modes in these diagrams and elsewhere in the paper because they are typically negligible with respect to the scalar contribution in models of single-field inflation featuring an enhanced scalar spectrum.

\section{CONSISTENCY RELATION}

\noindent The angular coefficients in eq.\,(\ref{eq:spherical_bessel}) can be obtained by calculating the correlator $\langle\delta_{\rm GW}({\bm x},{\bm q})\delta_{\rm GW}({\bm y},{\bm q})\rangle$ in Fourier space. Eq.\,(\ref{eq:gw_energy}) becomes, using eq.\,(\ref{eq:h_solution}),
\begin{widetext}
\begin{align*}
\rho_{\rm GW}(\tau,&{\bm x},{\bm q})
=
\frac{1}{16a^2}\frac{q^3}{2\pi^2}
\int\frac{d^3p}{(2\pi)^3}
\int\frac{d^3k}{(2\pi)^3}
\int\frac{d^3\ell}{(2\pi)^3}
\frac{(q^2-{\bm p}\cdot{\bm q})}{q^2|{\bm p}-{\bm q}|^2}
e^{i{\bm p}\cdot {\bm x}}
\mathcal{R}_{\bm k}\mathcal{R}_{{\bm p}-{\bm q}-{\bm k}}
\mathcal{R}_{\bm \ell}\mathcal{R}_{{\bm q}-{\bm \ell}}
\\&
{\rm Tr}\Big[{\bm e}^s({\bm p}-{\bm q})\cdot{\bm e}^t({\bm q})\Big]
\Big[{\bm k}\cdot {\bm e}^s({\bm p}-{\bm q})\cdot{\bm k}\Big]
\Big[{\bm \ell}\cdot {\bm e}^t({\bm q})\cdot{\bm \ell}\Big]
\overline{
I_{|{\bm p}-{\bm q}|}(k,|{\bm p}-{\bm q}-{\bm k}|)
I_q(\ell,|{\bm q}-{\bm \ell}|)
}.\numberthis
\label{eq:full_omega_detail}
\end{align*}
\end{widetext}
The dependence of $I$ on $\tau$ disappears after averaging and taking the late-time limit, so we have omitted it. To be clear, $\hat{f}\propto \Omega_{\rm GW}$ in eq.\,(\ref{eq:density_cont}) corresponds to the Gaussian result, obtained by taking the ensemble average of the above quantity and contracting $\langle\mathcal{R}^4\rangle\sim \langle\mathcal{R}^2\rangle^2$. Thus, $\delta_{\rm GW}$ can be found by considering all possible non-Gaussian contractions in this expression.

The calculation of the correlator $\langle\delta_{\rm GW}^2\rangle$ involves squaring this quantity and contracting $\langle\mathcal{R}^8\rangle$ according to eq.\,(\ref{eq:master-diagram}). Let us define the bispectra of $\mathcal{R}$ and $\varphi$ as
\begin{align}
\langle
\mathcal{R}_{\bm p}
\mathcal{R}_{\bm k}
\mathcal{R}_{\bm q}
\rangle
&=
B_\mathcal{R}({\bm p},{\bm k},{\bm q})(2\pi)^3\delta^3({\bm p}_{\rm total}).
\\
\langle
\varphi_{\bm p}
\varphi_{\bm k}
\varphi_{\bm q}
\rangle
&=
B_\varphi({\bm p},{\bm k},{\bm q})(2\pi)^3\delta^3({\bm p}_{\rm total}).
\end{align}
We also define
\begin{equation}
\langle
\mathcal{R}_{\bm p}
\mathcal{R}_{\bm k}
\mathcal{R}_{\bm q}
\mathcal{R}_{\bm \ell}
\rangle_s
=
S_\mathcal{R}({\bm p},{\bm k},{\bm q},{\bm \ell})
(2\pi)^3\delta^3({\bm p}_{\rm total}),
\end{equation}
where the $s$ subscript on the left-hand side denotes the fact that we consider only the tree-level $s$-channel diagram in the four-point function, which corresponds exactly to the diagrams in (\ref{eq:s_channel}) once the exchanged momentum ${\bm p}+{\bm k}=-({\bm q}+{\bm \ell})$ is taken to vanish. In the supplemental material we show that, in this limit, we have
\begin{equation}
\frac{S_\mathcal{R}({\bm p},{\bm q})}{|\varphi_{\rm L}|^2}
=
2\bigg[
4f_{\rm NL}|\varphi_p|^2
+
\frac{B_\varphi({\bm p})}{|\varphi_{\rm L}|^2}
\bigg]\cdot({\bm p}\leftrightarrow {\bm q}),
\label{eq:key_eq_s}
\end{equation}
where $S_\mathcal{R}$ has only two arguments due to taking the limit ${\bm p}+{\bm k}\rightarrow 0$ and $B_\varphi({\bm p})$ denotes the bispectrum in the squeezed limit, so that ${\bm q}=-{\bm p}$ and ${\bm k}\rightarrow 0$. The bispectrum for $\mathcal{R}$ in the squeezed limit is, after performing the gauge transformation,
\begin{equation}
\frac{B_\mathcal{R}({\bm p})}{|\varphi_{\rm L}|^2}
=
4f_{\rm NL}|\varphi_p|^2
+
\frac{B_\varphi({\bm p})}{|\varphi_{\rm L}|^2}
=
-\frac{d\log \mathcal{P}_\mathcal{R}}{d\log p},
\end{equation}
where the last equality comes from the consistency relation \cite{Maldacena:2002vr,Creminelli:2004yq}.

Squaring eq.\,(\ref{eq:full_omega_detail}) and using the above results, we find, after straightforward manipulation,
\begin{equation}
|\delta_{\rm GW}(q)|^2
=
\frac{\Omega_{\rm NG}(q)^2}{\Omega_{\rm GW}(q)^2}|\varphi_{\rm L}|^2,
\end{equation}
where the gravitational wave frequency spectrum is
\begin{align*}
\Omega_{\rm GW}(q)
&=
\frac{1}{12}\int\frac{d^3p}{(2\pi)^3}
\frac{1}{2\pi^2q}
\Big[{\bm p}\cdot{\bm e}^s({\bm q})\cdot{\bm p}\Big]^2
\\&\qquad\quad
\frac{q^2}{a^2H^2}\overline{I_q(p,|{\bm q}-{\bm p}|)^2}
|\varphi_{|{\bm q}-{\bm p}|}|^2|\varphi_p|^2
,\numberthis
\end{align*}
and $\Omega_{\rm NG}$ can be obtained by modulating the above expression with the tilt of the power spectrum, yielding eq.\,(\ref{eq:omega-ng-main}). Using the above results for $|\delta_{\rm GW}(q)|^2\propto \mathcal{P}_{\delta_{\rm GW}}$, together with the fact that $|\varphi_{\rm L}|^2\propto \mathcal{P}^{\rm L}_\mathcal{R}\sim 10^{-9}$ is approximately scale invariant (so that $\mathcal{P}_{\delta_{\rm GW}}$ does not depend on $k$) in eq.\,(\ref{eq:spherical_bessel}), we find eq.\,(\ref{eq:anisotropies}) \cite{Bartolo:2019zvb,Li:2023xtl}. The first term in this expression accounts for anisotropies in the gravitational wave production, and the second one for the anisotropies accumulated when the tensor modes propagate in an inhomogeneous Universe. The former arises because of inflationary non-Gaussianities and is related to the tilt of the scalar power spectrum, whereas the latter depends only on the tilt of the tensor spectrum. Eq.\,(\ref{eq:omega-ng-main}), viewed in conjunction with (\ref{eq:anisotropies}), is the main result of this paper.

\section{CONCLUSIONS}

\noindent We have shown that scalar-induced gravitational wave anisotropies in models of canonical single-field inflation with an enhanced scalar spectrum are entirely determined by the tilt of the scalar and tensor power spectra, which encode information about the anisotropies generated at the time of production, and those arising from propagation in an inhomogeneous Universe, respectively. These results are shown in eqs.\,(\ref{eq:anisotropies}) and (\ref{eq:omega-ng-main}). This consistency relation \footnote{Let us mention that a different consistency relation for anisotropies was also presented in \cite{Bartolo:2019yeu}, where it was shown that the three-point function of the density contrast $\delta_{\rm GW}$ is related to the tilt of the two-point function.} could be used as a smoking gun in determining whether a peak in the gravitational wave spectrum is induced by scalar modes or has a different physical origin. This result, valid for gravitational waves induced after the end of inflation, is quite general, since the only approximations involved in the derivation are neglecting loop corrections and ignoring the effect of tensor propagators in the diagrams. This is typically justified, for instance, in models of ultra-slow-roll inflation, where the first-order tensor power spectrum is negligible with respect to the induced one. In fact, it is not even necessary to specify the couplings in the cubic Lagrangian explicitly, see eq.\,(\ref{eq:cubic_lagrangian}).

Let us emphasize that although eq.\,(\ref{eq:anisotropies}) has been derived before, e.g.\,in \cite{Bartolo:2019zvb,Li:2023xtl} in the context of local non-Gaussianity or in \cite{Ruiz:2024weh} by keeping the leading self-interaction term in the inflaton Lagrangian, here we have shown for the first time that eq.\,(\ref{eq:omega-ng-main}) is a completely general result which applies to all models of canonical, single-field inflation with an enhanced scalar spectrum.

A few comments regarding future directions of research are in order. Let us start by noting that our results show that enhancing the anisotropies in this class of models is difficult, since the spectrum must change quite abruptly for this term to grow significantly. The anisotropies produced via this mechanism in fact tend to be quite small \cite{Ruiz:2024weh} and, as discussed in the Introduction, more work is required to determine the prospects for measuring them. Moreover, it has been argued that the squeezed limit of the scalar bispectrum leads to no deviations from Gaussianity in certain contexts \cite{Pajer:2013ana}. The results of \cite{Pajer:2013ana} cannot be applied directly to our calculation because we do not study the measurement of a squeezed three-point function directly, but rather its effect on the initial conditions of a different observable (namely, the two-point function $\langle\delta_{\rm GW}^2\rangle$). Nonetheless, assessing how these arguments impact the calculation presented here is an interesting prospect. Determining whether the relation presented here survives loop corrections and how it is affected by the propagation of tensor modes in the diagrams is also an interesting possibility.

\bigskip
\noindent {\bf Acknowledgments}. We are deeply grateful to Jes\'us Gamb\'in Egea for a thorough reading of the manuscript, as well as the calculation of the bispectrum in the squeezed limit \cite{to-appear}. We are also grateful to Guillermo Ballesteros for his many comments on the draft, and to Gabriele Franciolini for pointing out the arguments in \cite{Pajer:2013ana}. This work is supported by the Deutsche Forschungsgemeinschaft under Germany’s Excellence Strategy – EXC 2121 Quantum Universe – 390833306.

\bibliography{mybib}

\newpage

\section{SUPPLEMENTAL MATERIAL: \\ SQUEEZED CORRELATORS}

\noindent The key relation we must show in order to derive eq.\,(\ref{eq:omega-ng-main}) is eq.\,(\ref{eq:key_eq_s}). We do this by brute force calculation. The starting point for the proof is the third-order Lagrangian in the $\delta\phi$ gauge \cite{Maldacena:2002vr}
\begin{align*}
\mathcal{L}
&=
-c_1\varphi^3
-c_2\varphi\dot{\varphi}^2
-c_3\dot{\varphi}\partial_i\varphi\partial^i(\partial^{-2}\dot{\varphi})
-c_4\varphi\partial_i\varphi\partial^i\varphi
\\&
-c_5\varphi\partial_i\partial_j(\partial^{-2}\dot{\varphi})
\partial^i\partial^j(\partial^{-2}\dot{\varphi})
-\frac{d}{dt}(c_6\varphi^3),\numberthis
\label{eq:cubic_lagrangian}
\end{align*}
which contains every cubic interaction in canonical single-field inflation, including boundary terms. The coefficients are functions of the Hubble rate $H$ and the slow-roll parameters, though the expressions are irrelevant for our purposes. One can easily check that, up to cubic order, the interaction Hamiltonian is $\mathcal{H}=-\mathcal{L}$.

The tree-level bispectrum can be calculated using eq.\,(\ref{eq:in-in-master}). The result is \cite{to-appear}
\begin{align*}
B_\varphi({\bm p},{\bm q},{\bm k})
&=
-\int_{-\infty}^t dt'
2{\rm Im}\Big[\sum_n B_{(n)}(t,t'|{\bm p},{\bm q},{\bm k})
\\&
+
({\bm p}\leftrightarrow {\bm q},{\bm k})\Big],\numberthis
\end{align*}
where, omitting the arguments for readability,
\begin{align*}
B_{(1)}
&=
2c_1(t')
\Phi_p(t,t')
\Phi_q(t,t')
\Phi_k(t,t'),
\\
B_{(2)}
&=
2c_2(t')
\Phi_p(t,t')
\Psi_q(t,t')
\Psi_k(t,t'),
\\
B_{(3)}
&=
c_3(t')
\Psi_p(t,t')
\bigg[
\frac{{\bm q}\cdot{\bm k}}{k^2}
\Phi_q(t,t')
\Psi_k(t,t')
+({\bm q}\leftrightarrow {\bm k})
\bigg],
\\
B_{(4)}
&=
-2c_4(t')({\bm q}\cdot{\bm k})
\Phi_p(t,t')
\Phi_q(t,t')
\Phi_k(t,t'),
\\
B_{(5)}
&=
2c_5(t')\frac{({\bm q}\cdot{\bm k})^2}{q^2k^2}
\Phi_p(t,t')
\Psi_q(t,t')
\Psi_k(t,t').\numberthis
\end{align*}
In these expressions we have used the following shorthand quantities
\begin{equation}
\Phi_p(t,t')
\equiv
\varphi_p(t')\varphi_p^\star(t),
\quad
\Psi_p(t,t')
\equiv
\dot{\varphi}_p(t')\varphi_p^\star(t).
\end{equation}
Taking the squeezed limit $k\ll p=q$, we have, using the fact that $\varphi_k(t')\simeq \varphi_k(t)$ is frozen on superhorizon scales,
\begin{align*}
\frac{B_\varphi({\bm p})}{|\varphi_{\rm L}|^2}
&=
-2{\rm Im}\bigg\{
\int_{-\infty}^t dt'
\Big[6c_1(t')+2p^2c_4(t')\Big]\Phi_p(t,t')^2
\\&
+
\Big[2c_2(t')+2c_5(t')\Big]\Psi_p(t,t')^2
\bigg\}.\numberthis
\end{align*}

We now focus on the calculation of the $s$-channel correlator $S_\mathcal{R}$. The $c_6$ term does not contribute to the bispectrum, so we keep it throughout the calculation and explicitly check that it drops out of $S_\mathcal{R}$ in the limit in which the exchanged momentum vanishes. Using eq.\,(\ref{eq:in-in-master}), we can write
\begin{equation}
S_\varphi({\bm p},{\bm k},{\bm q},{\bm \ell})
=
\int_{-\infty}^t dt'
\int_{-\infty}^{t'} dt''
\sum_{ij}
S_{(i,j)},
\end{equation}
where the $(i,j)$ indices range from $1$ to $6$ and
\begin{align*}
S_{(i,j)}
&=
2{\rm Re}\Big\{
\Big[
C_{(i)}^+(t,t'|{\bm p},{\bm k})
-
C_{(i)}^-(t,t'|{\bm p},{\bm k})
\Big]
\\&\qquad
C_{(j)}^+(t,t''|{\bm q},{\bm \ell})^\star
\Big\}
+
({\bm p},{\bm k}\leftrightarrow {\bm q},{\bm \ell}).
\numberthis
\end{align*}

The $C_{(i)}^\pm$ functions can be found by straightforward calculation. We have (we omit the arguments for readability)
\begin{align*}
C_{(1)}^+
&=
6\big[c_1(t')+\dot{c}_6(t')\big]
\Phi_p(t,t')
\Phi_k(t,t')
\varphi_{|{\bm p}+{\bm k}|}(t'),
\\
C_{(2)}^+
&=
c_2(t')
\Big[
\Psi_p(t,t')
\Psi_k(t,t')
\varphi_{|{\bm p}+{\bm k}|}(t')
\\&+
2\Phi_p(t,t')
\Psi_k(t,t')
\dot{\varphi}_{|{\bm p}+{\bm k}|}(t')
+
({\bm p}\leftrightarrow{\bm k})
\Big],
\\
C_{(3)}^+
&=
\bigg\{
c_3(t')
\big[{\bm p}\cdot({\bm p}+{\bm k})\big]
\Psi_k(t,t')
\bigg[
\frac{1}{p^2}
\Psi_p(t,t')
\varphi_{|{\bm p}+{\bm k}|}(t')
\\&+
\frac{1}{|{\bm p}+{\bm k}|^2}
\Phi_p(t,t')
\dot{\varphi}_{|{\bm p}+{\bm k}|}(t')
\bigg]
+({\bm p}\leftrightarrow {\bm k})
\bigg\}
\\&
+c_3(t')
(
{\bm k}\cdot{\bm p}
)
\dot{\varphi}_{|{\bm p}+{\bm k}|}(t')
\bigg[
\frac{1}{k^2}
\Phi_p(t,t')
\Psi_k(t,t')
\\&
+\frac{1}{p^2}
\Psi_p(t,t')
\Phi_k(t,t')
\bigg],
\\
C_{(4)}^+
&=
-2c_4(t')(p^2+3{\bm k}\cdot {\bm p}+k^2)
\Phi_p(t,t')
\Phi_k(t,t')
\\&
\qquad \varphi_{|{\bm p}+{\bm k}|}(t'),
\\
C_{(5)}^+
&=
\bigg\{
2c_5(t')
\frac{(k^2+{\bm k}\cdot{\bm p})^2}{k^2|{\bm p}+{\bm k}|^2}
\Phi_p(t,t')
\Psi_k(t,t')
\dot{\varphi}_{|{\bm p}+{\bm k}|}(t')
\\&
+
({\bm p}\leftrightarrow{\bm k})
\bigg\}
+
2c_5(t')
\frac{({\bm k}\cdot{\bm p})^2}{p^2k^2}
\Psi_p(t,t')
\Psi_k(t,t')
\\&
\qquad \varphi_{|{\bm p}+{\bm k}|}(t')
,
\\
C_{(6)}^+
&=
3c_6(t')
\Big[
\Phi_p(t,t')
\Phi_k(t,t')
\dot{\varphi}_{|{\bm p}+{\bm k}|}(t')
\\&+
2\Phi_p(t,t')
\Psi_k(t,t')
\varphi_{|{\bm p}+{\bm k}|}(t')
+
({\bm p}\leftrightarrow{\bm k})
\Big].\numberthis
\end{align*}
The expressions for $C_{(i)}^-$ can be obtained from these by conjugating all fields with momentum $p$ and $k$ and leaving the ones with $|{\bm p}+{\bm k}|$ unchanged.

Upon taking the limit ${\bm p}+{\bm k}\rightarrow 0$, we obtain $C_{(3)}^+=0$ and
\begin{align*}
\frac{C_{(6)}^+}{12c_6(t')}
&=
\Phi_p(t,t')
\Psi_p(t,t')
\varphi_{\rm L},
\\
\frac{C_{(2)}^+}{2c_2(t')}
=
\frac{C_{(5)}^+}{2c_5(t')}
&=
\Psi_p(t,t')^2
\varphi_{\rm L},
\\
\frac{C_{(1)}^+}{6\big[c_1(t')+\dot{c}_6(t')\big]}
=
\frac{C_{(4)}^+}{2p^2c_4(t')}
&=
\Phi_p(t,t')^2
\varphi_{\rm L}.\numberthis
\end{align*}
Summing over all the terms we find
\begin{equation}
\frac{S_\varphi({\bm p},{\bm q})}{|\varphi_{\rm L}|^2}
=
2
\frac{B_\varphi({\bm p})}{|\varphi_{\rm L}|^2}
\frac{B_\varphi({\bm q})}{|\varphi_{\rm L}|^2}
+(\propto c_6).
\label{eq:ident_part}
\end{equation}
The last term, proportional to $c_6$, in fact vanishes upon integration. Let us first note that the time integrals can be written as
\begin{equation}
\int_{-\infty}^t
dt'
\int_{t'}^t
dt''
f(t',t'')
=
\int_{-\infty}^t
dt'
\int_{-\infty}^{t'}
dt''
f(t'',t'),
\label{eq:integral_swap}
\end{equation}
where $f$ is a generic function. We then have
\begin{align*}
(\propto c_6)
&=
12|\varphi_{\rm L}|^2
\int_{-\infty}^t
dt'
\int_{t'}^t
dt''
\\&
\Big\{
\Theta_p(t,t'')
\frac{d}{dt'}
2{\rm Im}\Big[
c_6(t')
\Phi_q(t,t')^2
\Big]
\\&
+3
\frac{d}{dt''}
2{\rm Im}
\Big[
c_6(t'')
\Phi_p(t,t'')^2
\Big]
\frac{d}{dt'}
2{\rm Im}
\Big[
c_6(t')
\Phi_q(t,t')^2
\Big]
\\&
+\frac{d}{dt''}
2{\rm Im}\Big[
c_6(t'')
\Phi_p(t,t'')^2
\Big]
\Theta_q(t,t')
+
({\bm p}\leftrightarrow {\bm q})
\Big\},\numberthis
\end{align*}
where
\begin{align*}
\Theta_p(t,t')
&=
2{\rm Im}
\Big\{
\big[3c_1(t')+c_4(t')p^2
\big]
\Phi_p(t,t')^2
\\&
+
\big[c_2(t')+c_5(t')\big]
\Psi_p(t,t')^2
\Big\}.\numberthis
\end{align*}
The first two terms vanish upon integration in $t'$. The evaluation of the integrand in the upper limit involves the imaginary part of an absolute, whereas the evaluation in the lower limit vanishes after taking into account the $i\epsilon$ prescription for subhorizon modes \cite{Weinberg:2005vy}. For the third term we can use eq.\,(\ref{eq:integral_swap}) again and integrate the total derivative in $t'$, which vanishes by the same argument.

The identity in (\ref{eq:ident_part}) corresponds to the middle diagram in eq.\,(\ref{eq:s_channel}). It is straightforward to account for the gauge transformation and show that the diagram on the left yields the $f_{\rm NL}^2|\varphi_p|^2|\varphi_q|^2$ factor in (\ref{eq:key_eq_s}), whereas the one on the right corresponds to the $f_{\rm NL}|\varphi_p|^2B_\varphi({\bm q})$ factors.

\end{document}